\documentclass[letterpaper, english, reprint, citeautoscript, groupedaddress, superscriptaddress] {revtex4-1}

\usepackage[T1]{fontenc}
\usepackage[latin9]{inputenc}
\usepackage{babel}
\usepackage{textcomp}
\usepackage[Symbol]{upgreek}
\usepackage{calc}
\usepackage{amssymb}
\usepackage{amsmath}
\usepackage{graphicx}
\usepackage{adjustbox}
\usepackage{xfrac}
\usepackage{bm}
\usepackage{dsfont}
\usepackage[colorlinks=true,linkcolor=black, citecolor=blue, urlcolor=blue, unicode=true]{hyperref}
\usepackage{xspace}

\newcommand{\Hintc}[1]{\ensuremath{\mu_0 H^{\text{int}}_{\text{c}#1}}}
\newcommand{\bHint}{\ensuremath{\mu_0\bm{H}_{\text{int}}}\xspace}
\newcommand{\Hint}{\ensuremath{\mu_0 H_{\text{int}}}\xspace}
\newcommand{\SqEH}{\ensuremath{S\left(\bm{q},E,\bHint\right)}\xspace}
\newcommand{\Tc}{\ensuremath{T_{\text{c}}}\xspace}
\newcommand{\kh}{\ensuremath{k_{\text{h}}}\xspace}
\newcommand{\Ei}{\ensuremath{E_{\text{i}}}\xspace}
\newcommand{\Ef}{\ensuremath{E_{\text{f}}}\xspace}
\newcommand{\bq}{\ensuremath{\bm{q}}\xspace}
\newcommand{\bqpar}{\ensuremath{\bq_{\parallel}}\xspace}
\newcommand{\qpar}{\ensuremath{q_{\parallel}}\xspace}
\newcommand{\qper}{\ensuremath{q_{\perp}}\xspace}
\newcommand{\bqper}{\ensuremath{\bq_{\perp}}\xspace}


\begin{document}
\preprint{}

\title{Non-reciprocal magnons in non-centrosymmetric MnSi}

\newcommand{\tum}{Physik-Department, Technische Universit\"at M\"unchen (TUM), James-Franck-Str. 1, 85748 Garching, Germany}
\newcommand{\mlz}{Heinz-Maier-Leibnitz-Zentrum (MLZ), Technische Universit\"at M\"unchen (TUM), Lichtenbergstr. 1, 85747 Garching, Germany}
\newcommand{\lns}{Laboratory for Neutron Scattering and Imaging, Paul Scherrer Institut (PSI), CH-5232 Villigen, Switzerland}
\newcommand{\qm}{Laboratory for Quantum Magnetism, \'Ecole Polytechnique F\'ed\'erale de Lausanne, CH-1015 Lausanne, Switzerland}
\newcommand{\juelichill}{J\"ulich Centre for Neutron Science (JCNS), Forschungszentrum J\"ulich GmbH, Outstation at Institut Laue-Langevin, Bo\^ite Postale 156, 38042 Grenoble Cedex 9, France}
\newcommand{\cologne}{Institut f\"ur Theoretische Physik, Universit\"at zu K\"oln, Z\"ulpicher Str. 77a, 50937 K\"oln, Germany}
\newcommand{\dresden}{Institut f\"{u}r Theoretische Physik, Technische Universit\"{a}t Dresden, D-01062 Dresden, Germany}
\newcommand{\ill}{Institut Laue-Langevin (ILL), 71 avenue des Martyrs, 38000 Grenoble, France}

\author{T. Weber}
\email[Correspondence: ]{tweber@ill.fr}
\affiliation{\ill}

\author{J. Waizner}
\affiliation{\cologne}

\author{G. S. Tucker}
\affiliation{\lns}
\affiliation{\qm}

\author{L. Beddrich}
\affiliation{\mlz}
\affiliation{\tum}

\author{M. Skoulatos}
\affiliation{\mlz}
\affiliation{\tum}

\author{R. Georgii}
\affiliation{\mlz}
\affiliation{\tum}

\author{A. Bauer}
\affiliation{\tum}

\author{C. Pfleiderer}
\affiliation{\tum}

\author{M. Garst}
\affiliation{\dresden}

\author{P. B\"oni}
\affiliation{\tum}

\date{\today}

\begin{abstract}
	Using two cold-neutron triple-axis spectrometers we have succeeded in fully mapping out the field-dependent evolution of the non-reciprocal magnon dispersion relations in all magnetic phases of MnSi.
	The non-reciprocal nature of the dispersion manifests itself in a full asymmetry (non-reciprocity) of the dynamical structure factor \SqEH with respect to flipping either the direction of the applied magnetic field \bHint, the reduced momentum transfer ${\bm q}$, or the energy transfer $E$.

	\vspace{0.2cm}
	\noindent This is a pre-print of our conference paper \cite{weber2018non}, which is available online at \url{https://doi.org/10.1063/1.5041036}. Except where otherwise stated, e.g. for Figs. \ref{fig:fp} and \ref{fig:coni}, the paper is licensed under CC BY 4.0 (\url{http://creativecommons.org/licenses/by/4.0/}).
\end{abstract}

\maketitle

\section{ \label{sec:intro} Introduction }

The chiral itinerant-electron magnet MnSi crystallizes in the non-centrosymmetric space group $\mathrm{P2_13}$.
A lack of inversion symmetry enforces a Dzyaloshinskii-Moriya interaction and gives rise to the rich magnetic phase diagram of MnSi which features \cite{Muehl09}:
a helical arrangement of the Mn electron spins with a pitch of $\kh=0.036$ \AA$^{-1}$ for applied magnetic fields and temperatures below the critical values $\Hintc{1}=0.1$ T and $\Tc=29.5$ K, respectively \cite{Jano10};
a conical spin phase is found for fields above \Hintc{1} and below a second critical field $\Hintc{2}=0.55$ T;
above \Hintc{2} the spins align along the applied field direction forming the field-polarized phase;
lastly, and most interestingly, a small region at the border of the conical phase has been found to contain a skyrmion phase where the Mn magnetic moments align in a vortex-like fashion \cite{Muehl09}.

In the early 1980s, G.~Shirane, et al. \cite{Shirane83}, found the first clues to asymmetric scattering phenomena in MnSi caused by its non-centrosymmetric unit cell.
They discovered that the helical arrangement of the Mn moments is single-handed, i.e., either fully left- or fully right-handed.
As a result, they deduced that the dispersion in the field-polarized phase must be asymmetric with respect to the spin-flip channels.
Recently, the non-reciprocal behavior of the field-polarized state was further investigated by S.~Grigoriev, et al. \cite{Grigoriev15}, as well as T.~J.~Sato, et al. \cite{Sato16}, confirming that the dispersion in that region takes the form of asymmetric non-centered parabolic branches as shown in the right-hand panel of Fig.~\ref{fig:disp}.
In the paramagnetic phase above \Tc, B.~ Roessli, et al. \cite{Roessli2002}, found asymmetric contributions to critical scattering.
Janoschek, et al. \cite{JanoSkyrmi}, were the first to conduct investigations into the magnon dynamics of the skyrmion phase.

This contribution serves a twofold purpose: we first give a concise summary of our recent work on the non-reciprocal dynamics in the conical and field-polarized phases of MnSi \cite{Weber2017Field} and hereby also present new data;
secondly we show our very recent \cite{SkxAsym18} results which prove that the non-reciprocal phenomena of the conical phase persist into the elusive skyrmion phase.

\section{ \label{sec:exp} Experimental setups }

We employed the cold-neutron triple-axis spectrometers MIRA \cite{MIRAnew} at the Heinz Maier-Leibnitz Zentrum (MLZ) and TASP \cite{TASP96, Semadeni01} at the Paul Scherrer Institut (PSI) for our experiments in the helimagnetic, conical, field-polarized, and skyrmion phases of MnSi.
The results of the measurements for each phase are discussed in the following sections.

Experiments at MLZ and PSI used a vertically focussing monochromator, flat analyzer, and symmetric collimation before and after the sample to improve the instrumental resolution.
All experiments made use of a cooled beryllium filter, which does not transmit neutrons with energies above $5$ meV, to remove higher-order neutrons.
MIRA at MLZ was configured with 30' collimation, the filter before the sample, and fixed incident neutron energies between $3.19\leq\Ei\leq4.06$ meV.
TASP at PSI was configured with 40' collimation, the filter after the sample, and fixed final neutron energies between $4.06\leq\Ef\leq4.66$ meV.

All measurements were conducted around the $\bm{G} = \left( 110 \right)$ nuclear Bragg reflection of a cylindrical MnSi single crystal ($3$ cm by $1$ cm diameter) and corrected for demagnetization effects due to the sample geometry \cite{Sato89}.
Throughout this paper we utilize the reduced momentum transfer notation $\bq = \bm{Q}-\bm{G}$ such that \bq is always within the first Brillouin zone of the nuclear structure.
The individual measurements were performed with $\bq \parallel \bHint$, which is also indicated by the shorthand notation \bqpar. The reduced momentum transfer perpendicular to the field, $\bqper{}$, was zero for all scans in this study.
Most measurements were performed at $T=20$ K with the exception of those in the skyrmion phase where both temperature and field were scanned to yield optimal intensity on the magnetic skyrmion satellites.

\section{ \label{sec:fp} Field-polarized phase }

\begin{figure}
\includegraphics[width=1\columnwidth]{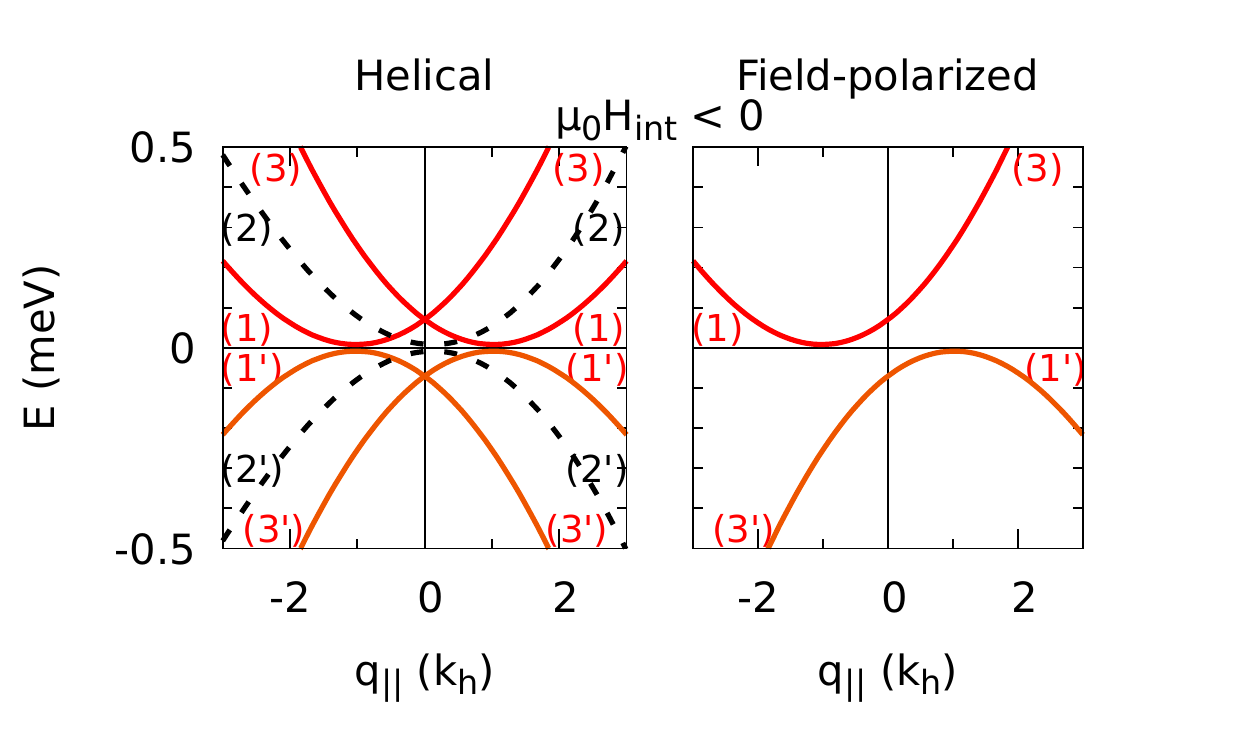}
\caption{Left: The dispersion in the helimagnetic and in the conical phases consists of three parabolic branches which are centered around the nuclear Bragg peak (dashed black) and the two magnetic satellite Bragg peaks (orange and red lines), respectively ($\bqper{}=0$).
Right: In the field-polarized phase, the nuclear-centered and one of the satellite-centered dispersion branches vanish depending on the sign of the magnetic field.}
\label{fig:disp}
\end{figure}

\begin{figure*}[ht]
\includegraphics[width=0.75\textwidth]{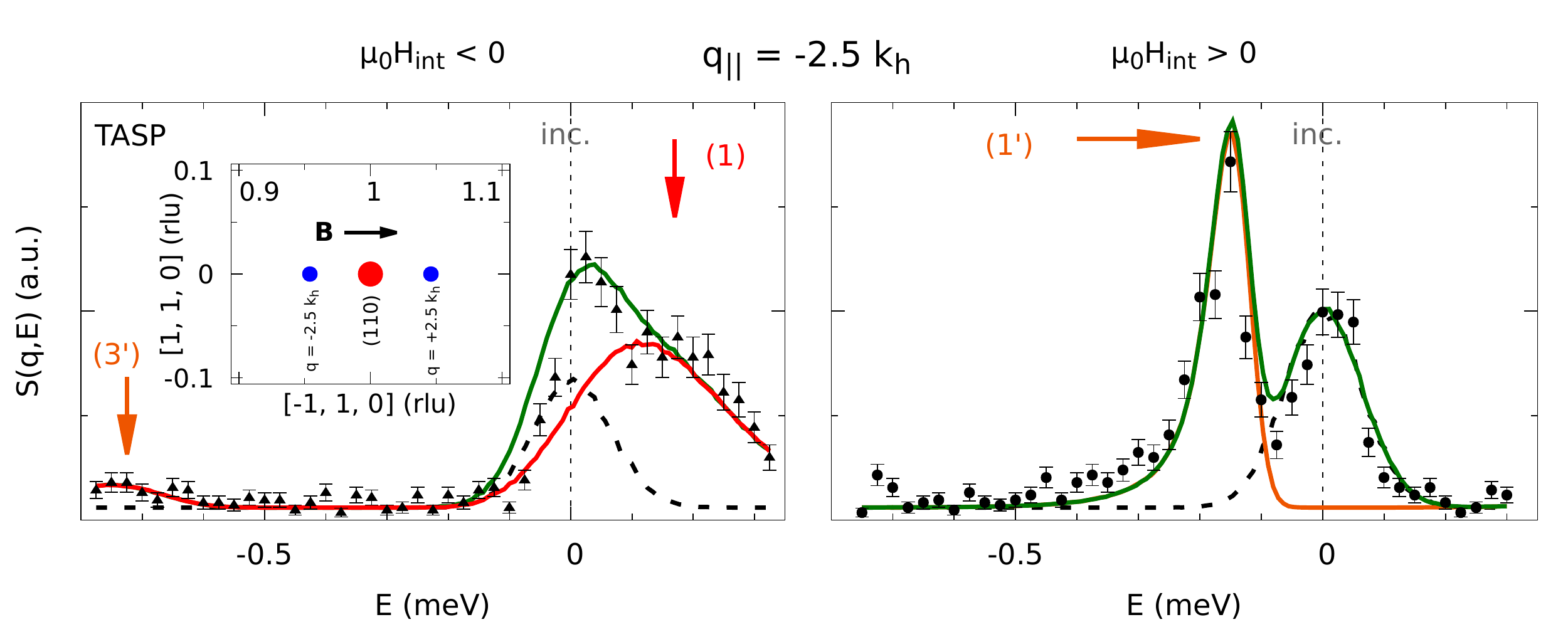}
\caption{Non-reciprocal magnons in the field-polarized ferromagnetic phase ($T = 20\ \mathrm{K}$, $\Hint = 610\ \mathrm{mT}$) \cite{Weber2017Field}.
	Left: The magnon (1) is created with a lower absolute energy than it is annihilated (3').
	The labels refer to the three parabolic branches in the helical phase, see Fig. \ref{fig:disp}.
	Right: Upon reversing the field \bHint, the magnon is now annihilated with a lower absolute energy than it is created (not shown).
	The lines are an instrumental resolution-convolution \cite{Takin2016, Takin2017} of the theoretical model \cite{Garst2017}.
	Figure reproduced with permission from Phys. Rev. B 97, 224403 (2018). Copyright 2018 American Physical Society. }
\label{fig:fp}
\end{figure*}

Our investigation of the field-polarized magnons \cite{Weber2017Field} comprised constant-\bq scans with $\bHint \parallel \left[ \overline{1}10 \right]$.
Here, the Mn spins fully align along the field and the magnetic structure becomes commensurate.
Nevertheless, the dispersion branches (Fig.~\ref{fig:disp}) are centered around the positions $\qpar=\pm\kh$ where in the helical and conical phase the incommensurate helimagnetic satellite reflections are located.
It is striking that the dispersion branches are asymmetric and centered around either $\qpar = +\kh$ or $\qpar = -\kh$ depending on whether the magnon is created or annihilated.
This behavior can be seen in the different absolute energies of the (1) and (3') peak in the left panel of Fig.~\ref{fig:fp}.
Reversing the field polarity (right panel of Fig.~\ref{fig:fp}) flips the center of the magnon creation ($E>0$) and annihilation ($E<0$) parabolic dispersion branches to the other satellite position.
Note that in Figs.~\ref{fig:disp} and \ref{fig:fp} both dispersion branches centered around the magnetic satellites at $\qpar = \pm\kh$ are labelled both (1) and (3), with the value of the label corresponding to the magnitude of the energy of the dispersion near $\kh$ or $-\kh$.
Furthermore, magnon creation branches are labelled with arabic numerals [e.g., (1)] and magnon annihilation by primed arabic numerals [e.g., (1')].

\section{ \label{sec:coni} Helical and conical phase }

\begin{figure*}[ht]
\includegraphics[width=0.75\textwidth]{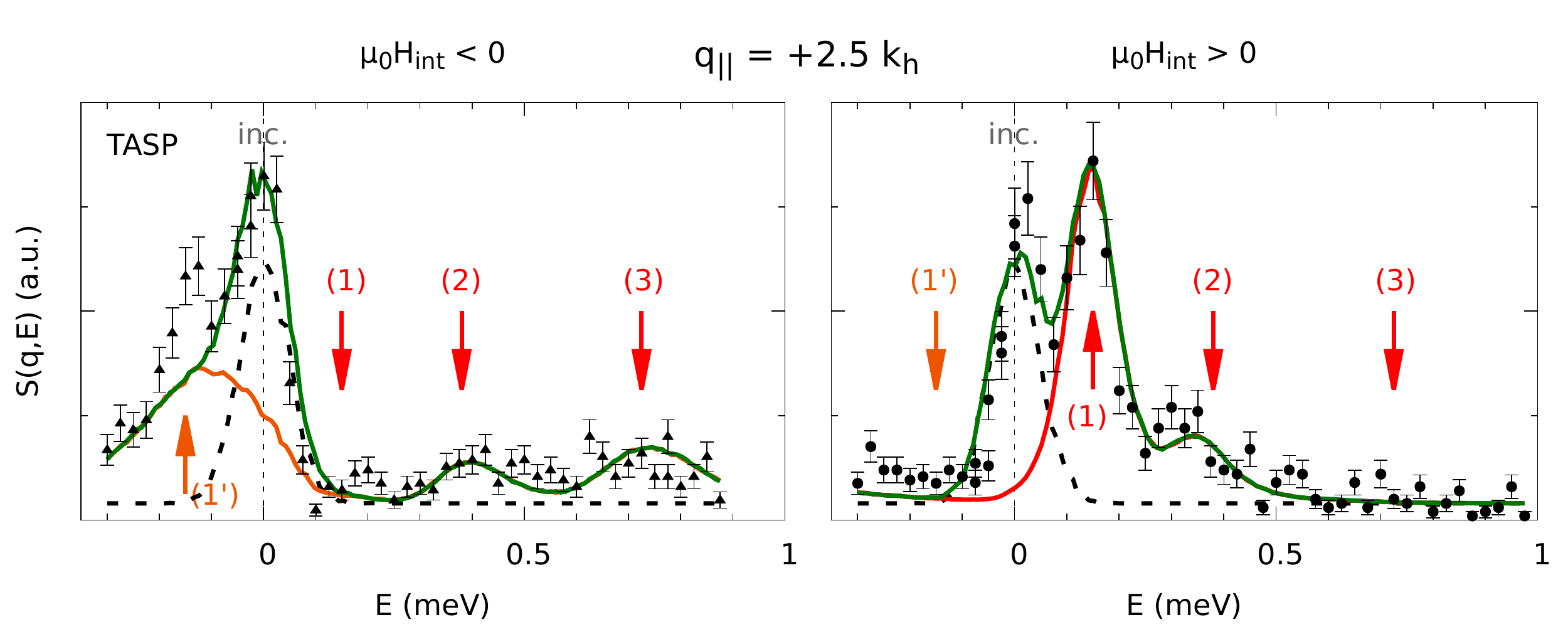}
\caption{Non-reciprocal magnons in the conical phase ($T = 20\ \mathrm{K}$, $\Hint = 440\ \mathrm{mT}$) \cite{Weber2017Field}.
	In difference to the field-polarized phase, all three parabolic branches (see Fig. \ref{fig:disp}) are theoretically allowed.
	The asymmetry in the dispersion is solely created by a field \bHint, momentum \bq, and energy $E$ dependent asymmetric distribution of the spectral weights.
	The two panels show the same scan with only the polarity of the field \bHint reversed.
	The lines are an instrumental resolution-convolution \cite{Takin2016, Takin2017} of the theoretical model \cite{Garst2017}.
	Please note that the mode (1') in the left panel is larger than what is expected by the convolution.
	Figure reproduced with permission from Phys. Rev. B 97, 224403 (2018). Copyright 2018 American Physical Society. }
\label{fig:coni}
\end{figure*}

Unlike the field-polarized phase, the helimagnetic \cite{Jano10, Kugler15} and conical \cite{Weber2017Field} phases are both marked by a symmetric dispersion relation, which consists of three parabolic branches of non-zero spectral weight (left panel of Fig. \ref{fig:disp}) for zero momentum transfer perpendicular to the helix propagation direction, $\qper=0$.
For $\qper\neq 0$, the small magnitude of \kh causes a pronounced backfolding of the magnon dispersion branches and an ensuing band-like dispersion \cite{Jano10, Kugler15, Garst2017}.
The three magnon branches at $\qper=0$ are centered around $\qpar=\pm\kh$ and $\qpar=0$, respectively.
For finite fields, the spectra nevertheless show the same asymmetries as in the field-polarized phase, only here the effect is purely due to an ever increasing asymmetry in the distribution of the spectral weights with increasing field, with the scattering geometry remaining the same as in the field-polarized measurements (Sec. \ref{sec:fp}).
A flipping of the spectral weights between the $\qpar= +\kh$ and the $\qpar= -\kh$ branches can be observed upon reversing the field polarity, see Fig. \ref{fig:coni}.
Here, the labels (1) and (3) refer to the dispersion branches which are centered around the magnetic satellites and (2) to the branch centered around the nuclear Bragg reflection.
For increasing fields, the branch centered at $q = 0$ loses all spectral weight and is forbidden in the fully field-polarized phase \cite{Weber2017Field}.

\section{ \label{sec:skx} Skyrmion phase }

\begin{figure*}[ht]
\includegraphics[width=0.75\textwidth]{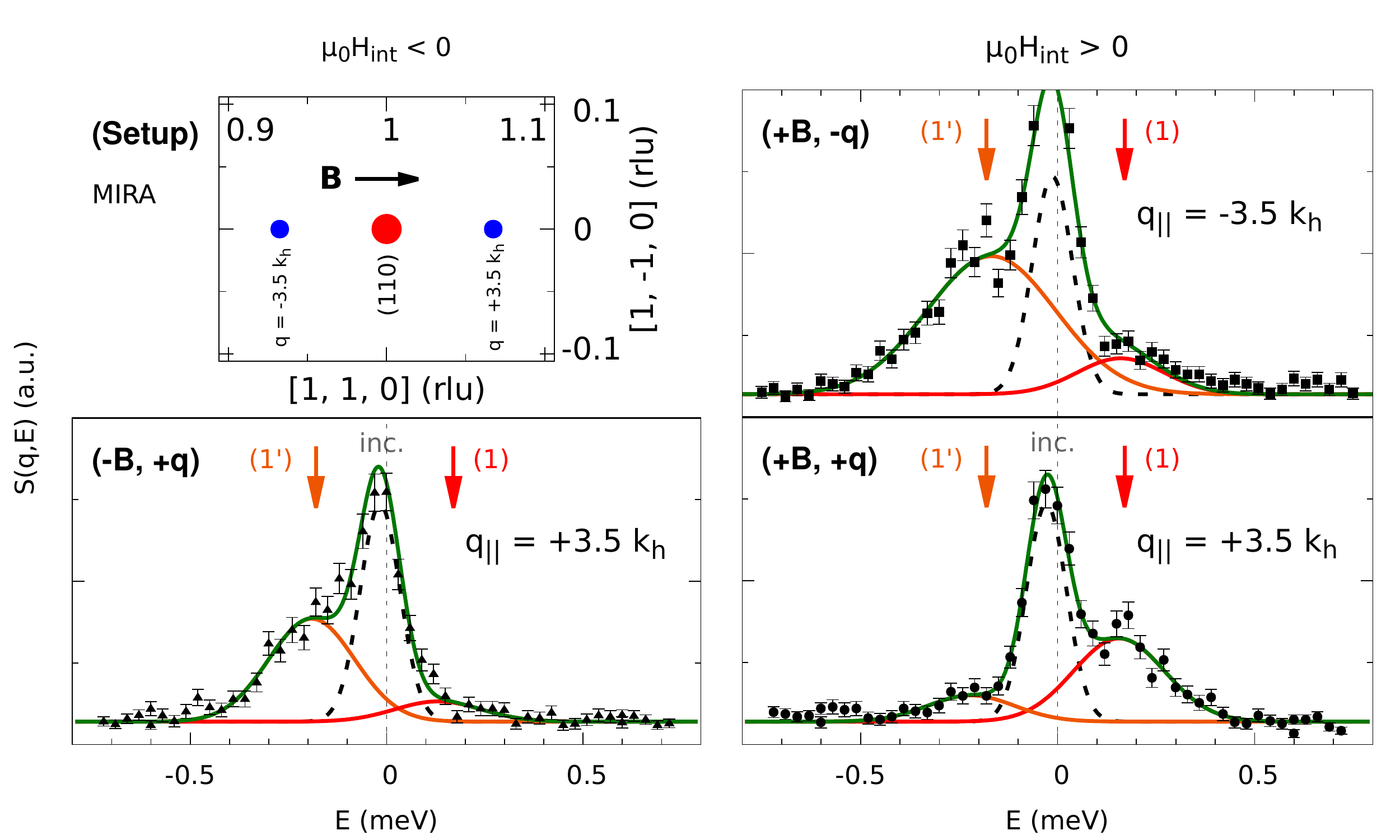}
\caption{Our very recent measurements of the non-reciprocal magnons in the skyrmion phase \cite{SkxAsym18} ($T = 28.5\ \mathrm{K}$, $\Hint= 150\ \mathrm{mT}$) found the same asymmetries upon reversal of either $E$, \bq or \bHint as in the conical and field-polarized phases.
	Furthermore, the discernible magnons have a lower stiffness than the magnons in the other phases, consistent with the symmetric dispersion \cite{JanoSkyrmi}.
	The lines are Gaussian fits and merely serve as guides to the eye. }
\label{fig:skx}
\end{figure*}

In order to investigate the skyrmion phase we performed constant-\bq scans with $\bHint\parallel \left[ 110 \right]$.
For this configuration, which is depicted in the top-left panel of Fig. \ref{fig:skx}, the skyrmion plane aligns perpendicular to the $(hk0)$ scattering plane with its normal vector parallel to \bHint.
For $\bq\parallel\bHint$ the dynamical structure factor \SqEH shows the same full non-reciprocity in all three variables \bq, $E$, and \bHint as in the conical and the field-polarized phases.
In Fig. \ref{fig:skx} we reversed the sign of the momentum transfer \bq (top-right panel) and the field (bottom-left) while keeping their absolute values, $q$ and \Hint, and all other measurement conditions the same as in the bottom-right panel.
Similar to the conical and field-polarized phases, the pronounced excitation labeled with (1) at positive energy transfers in the bottom-right panel $\left(+\bm{B},+\bq\right)$ reappears as excitation (1') at negative energy transfers in the top-right $\left(+\bm{B},-\bq\right)$ and bottom-left $\left(-\bm{B},+\bq\right)$ panels.
Only a sign change of two of the dependent variables leads back to the original dynamical structure factor.

\section{ \label{sec:concl} Conclusion}
Non-reciprocity of the magnetic fluctuations in the paramagnetic phase \cite{Roessli2002} as well as of the magnon dispersion in the field-polarized phase \cite{Shirane83, Grigoriev15, Sato16} of MnSi had been demonstrated previously.
In this work we have successfully expanded on those studies to identify non-reciprocity in the excitation spectra of all ordered magnetic phases of MnSi, thereby amending and concluding our prior work \cite{Weber2017Field}.
The recent availability of comprehensive time-of-flight and triple-axis data and novel theoretical models within our large collaboration provides a clear goal to expand on the current work in the near future \cite{fobes2018spin, SkxAsym18}.

\begin{acknowledgments}
This work is based upon experiments performed at the MIRA instrument operated by FRM II at the Heinz Maier-Leibnitz-Zentrum (MLZ), Garching, Germany and on experiments performed at the TASP instrument at the Swiss spallation neutron source SINQ, Paul Scherrer Institute (PSI), Villigen, Switzerland.
This work was part of the Ph.D. thesis of T.W. and was supported by the DFG under GE 971/5-1.
M.G. is supported by the DFG via SFB 1143 ``Correlated Magnetism: From Frustration to Topology'' and grant GA 1072/5-1. A.B. and C.P. gratefully acknowledge financial support through DFG TRR80 (project E1) and ERC Advanced Grant 291079 (TOPFIT).
We thank R. Schwikowski, J. Frank and M. Bartkowiak for technical support.
\end{acknowledgments}

\end{document}